\documentclass[preprint,prb, superscriptaddress,showpacs,nofootinbib,amsmath,amssymb,
aps,longbibliography]{revtex4-2}
\usepackage{framed}
\usepackage{graphicx}
\usepackage{dcolumn}
\usepackage{amsmath}
\usepackage{amssymb}
\usepackage{color}
\usepackage{subfigure,amsmath,verbatim,moreverb,bm}

\def\bef{\begin{framed}}
\def\eef{\end{framed}}
\def\be{\begin{equation}}
\def\ee{\end{equation}}
\def\ber{\begin{eqnarray}}
\def\eer{\end{eqnarray}}

\def\nablav{{\boldmath{\nabla}}}

\def\nuv{\mbox{\boldmath $\nu$}}

\def\rv{{\bf r}}

\def\vv{{\bf v}}

\def\jv{{\bf j}}

\def\Av{{\bf A}}
\def\Bv{{\bf B}}

\def\vv{{\bf v}}
\def\nn{\nonumber}

\begin{document}
\title{Current density functional theory in the age of generalized Kohn-Sham theories}
\author{Giovanni Vignale}
\email{vignaleg@missouri.edu}
\date{\today}

\begin{abstract}
I review the evolution of electronic non-relativistic current density functional theory (CDFT) from its original formulation in the late 1980's to present day's formulations, which rely on the  generalized Kohn-Sham theory.  Axel Becke's outstanding contributions are highlighted throughout. I propose a  form of the generalized Kohn-Sham equation that combines the exchange-correlation vector potential of the original formulation with an  ``effective mass term" -- the latter following from the inclusion of Becke's kinetic energy density as a basic input of the exchange-correlation energy functional.

\end{abstract}
\maketitle

\newpage
\section{Introduction}\label{Introduction}
I am honored to be invited to contribute to this Festschrift  honoring Axel Becke's outstanding contributions to density functional theory.  I will say immediately that my direct interaction with Axel was minimal -- in fact, I do not remember having any in-depth scientific discussion with him.  I attended  several conferences where he was an invited speaker, and read some of his papers, but missed the opportunity to develop a personal relation.  Our styles were very different. Axel was a great creator of density functional approximations: he was interested in functionals that work in practice. In contrast, I was a   formalist, by which I mean my main interest was  getting the right equations, not  solving them.  
Despite this, our paths crossed one beautiful day of June of 2002 when we were both attending the ``Walter Kohn Symposium"  organized by Dennis Salahub at the National Research Council in Ottawa.
In the symposium Axel presented a new kind of current density functional theory (CDFT) which, he declared, solved the long-standing problem of calculating the ground state energies of open shell atoms\cite{Becke2002}. Open shell atoms have degenerate ground states, which may be superimposed in different ways to generate states with different densities and orbital current densities: but all of them must have exactly  the same energy. It is not difficult to see why this creates a nightmare scenario for any density functional approximation.  The reason why his talk touched a nerve, though, was Axel's claim that he had solved the problem by including the current density dependence of the energy functional.   At that time I owed much of my modest scientific reputation to a couple of papers  I had published  some fifteen years earlier with the late Mark Rasolt (hereafter referred to as ``VR" theory), in which we formulated the current and spin density functional theory for electronic systems at high magnetic field\cite{VR1987,VR1988}.  In those papers I prided myself to have established that  (i)  a Hohenberg-Kohn-like theorem exists for the paramagnetic current density, $\jv_p(\rv)$, but not for the physical current density, $\jv(\rv)$ -- the two currents differing  by the diamagnetic term and (ii)  the gauge invariance of the exchange-correlation energy functional, $E_{xc}$,  leaves us with no choice but stipulating that  $E_{xc}$ must be a functional of the gauge-invariant combination $\nuv(\rv)=\nablav_\rv \times \frac{\jv_p(\rv)}{n(\rv)}$ (where $n(\rv)$ is the particle density) -- a quantity we dubbed ``vorticity".  Following the favorable reception of these papers, I had come to think of myself as an ``expert" in the niche area of current-dependent functionals. With my student Pawel Skudlarski, I had  even proposed an approximate functional\cite{Skudlarski93,Skudlarski93-E} for three dimensional electronic systems in a magnetic field.  I had  applied the theory with some success to fashionable two-dimensional systems like quantum dots\cite{Ferconi94}, quantum Hall edge states\cite{EdgeStates95}, and Wigner crystal at high magnetic field\cite{WignerCrystal93}.  
Now  came Axel Becke with his open shell atoms at zero magnetic field and essentially declared that what I had been saying for years was wrong: his functional was not a functional of the vorticity, it contained the paramagnetic current explicitly, and yet it was gauge-invariant, and solved a real-world problem!

The day after his presentation we had a brief chat during the coffee break. I must have said something to the effect that his calculations still allowed  energy differences to exist between  states that should have been  exactly degenerate. That did not seem to bother him in the least. He was very excited and visibly proud of the fact that his functional had tremendously reduced those energy differences, squeezing them down to values below the threshold of ``chemical accuracy".  But the thing that really bothered me as a formalist was that he had been able to  bypass my ``baby", i.e., the vorticity variable, without compromising the gauge-invariance of the functional.  Did that mean that I had been wrong all along?

The answer to this question is both a no and a yes, and in the rest of this paper I will try to explain how and why.
In brief, the analysis that led to the choice of the vorticity as the basic variable for the $E_{xc}$ functional was firmly grounded in the conventional Kohn-Sham framework in which the densities are the basic variables and the Kohn-Sham orbitals are just an auxiliary device to represent the densities.  But, at the time of my 2002 meeting with Axel Becke a crucial paradigm shift had already occurred from the conventional Kohn-Sham (KS) framework to what is now called the generalized Kohn-Sham (GKS) framework~\cite{Seidl96}. In this new framework the Kohn-Sham orbitals  effectively become  ``basic variables", in the sense that the self-consistent equations are obtained by zeroing the first order variation of the energy caused by an infinitesimal variation of the orbitals, not the densities: in other words, the variation of the orbitals determines the variation of the densities, not the other way around.  It turns out that in the GKS framework the vorticity  is no longer a necessary ingredient of a gauge-invariant $E_{xc}$ functional. Not only that:  the generalized Kohn-Sham equations that emerge in this  approximate framework are significantly different from the ``classic" ones. The most striking difference, as will be shown below, is that the xc vector potential, $\Av_{xc}(\rv)$,  -- the most distinctive signature of the VR-CDFT -- is no longer present in the equations: it is superseded   by an effective mass term which is quadratic, not linear, in the relative momentum operator $-i\nablav - \frac{\jv_p}{n}$.  Furthermore, the GKS equations determine the full (physical) current density directly, whereas the original VR equations directly determined only the paramagnetic current density. 

Axel Becke was of course very well aware of this paradigm shift,  being himself one of its main promoters. For me acceptance came much more slowly and against great resistance.  Throughout the first decade of 2000 the problem of degenerate states of open-shell atoms continued to attract considerable attention and was regularly tackled with the help of orbital-dependent functionals~\cite{Pittalis2006-1,Pittalis2006-2,Pittalis2007} usually within the framework of the ``optimized effective potential" approach~\cite{OEP92}.  Approximate functionals of the metaGGA family were found to require the inclusion of the current density to ensure  gauge-invariance. Particularly notable for the present discussion are the papers by J. Tao and John Perdew\cite{Tao2005,TaoPerdew2005} who attempted to reconcile the metaGGA xc functional with the original vorticity functional of VR (more about this later). But I suspect that these authors were still under the sway of the KS framework, because  they did not write down, much less tried to solve, the GKS equations that naturally follow from the current-dependent meta-GGA. Had they done so, they would have exposed the conceptual gap that separates the VR theory from the current-dependent meta-GGA. In Section~\ref{GeneralizedCDFT} I try to repair this omission.

For me the turning point came in 2017, when my long-term collaborators Stefano Pittalis and Florian Eich asked me to join them in an  paper\cite{Pittalis17} which defines the U(1) and SU(2) gauge-invariant building blocks of approximate density functionals of the currents and the spin currents. The influence of Axel Becke~\cite{Becke88,BR89} on this paper is impossible to overestimate. The transition to GKS at this point was complete.  Recent developments of spin current DFT~\cite{Desmarais2024,Desmarais2024-1,Desmarais2024-2,Desmarais2025,Desmarais2026} heavily rely on such building blocks. Vorticity is, apparently, sidelined, having in the meanwhile failed  some critical tests on small atomic systems\cite{ZhuTrickey2006} and struggled to incorporate formal complexities~\cite{Abedinpour2010}. But is it really so?  In the last section of this paper, after explaining the differences between the KS and the GKS formulations of CDFT, I will discuss the possibility of a  ``peaceful coexistence" between the two.

\section{``Classic" CDFT}\label{ClassicCDFT}

The development of a non-relativistic CDFT was strongly motivated by spectacular advances in the theory of interacting electrons at high magnetic field, such as, for example, the discovery of the fractional quantum Hall effect.
We quickly settled on the  paramagnetic current density, $\jv_p(\rv)$,  as the natural local density to which the vector potential would couple.  $\jv_p(\rv)$  is the density of canonical momentum divided by the electron mass and is  represented in terms of KS orbitals as follows
\be\label{ParamagneticCurrent}
\jv_p(\rv)=\frac{1}{2i}\sum_i\left\{\psi^*_i(\rv)[\nablav_\rv\psi(\rv)]-[\nablav_\rv \psi^*_i(\rv)]\psi(\rv)\right\}\,,
\ee
where the sum runs over occupied orbitals (we use atomic units, i.e., $e=\hbar=m=1$). On the other hand, the physical current density, $\jv(\rv)$, is constructed from the physical velocity operator (commutator of the position operator with the Hamiltonian)  and has the expression
\be\label{PhysicalCurrent}
\jv(\rv)= \jv_p(\rv)+n(\rv)\Av(\rv)
\ee
where 
\be\label{Density}
n(\rv)=\sum_i|\psi_i(\rv)|^2
\ee
is the particle density and $\Av(\rv)$ is the external vector potential, related to the magnetic field by $\Bv(\rv)=\nablav_\rv \times\Av(\rv)$. The paramagnetic current density, being independent of the external vector potential, allowed us to establish a natural separation of the energy into an ``internal" part (a universal functional of $n$ and $\jv_p$, independent of the potentials) and an ``external" part in which external potentials $V(\rv)$ and $\Av(\rv)$ couple to the local density and current density respectively: this separation  lies at the heart of the Hohenberg-Kohn theory.  We were also aware of the fact that the physical current density would vanish in a homogeneous electron gas in a uniform magnetic field, frustrating any attempts to formulate a local density approximation in terms of the physical current.  Although we missed  some subtle points, such as the possibility that different pairs of scalar and vector potentials  produce the same ground state~\cite{Capelle2002}, we were able to prove a Hohenberg-Kohn-like theorem for $n$ and $\jv_p$, i.e., we showed the existence of a unique map from paramagnetic current and density to ground state wave functions. Remarkably, such a theorem cannot be established in terms of the physical current density~\cite{Tellgren2012}.

At this point, according to the original Hohenberg-Kohn program, we should have come up with some approximation for the full energy functional and minimized it with respect to the  densities $n(\rv)$ and $\jv_p(\rv)$ at constant external potentials $V(\rv)$ and $\Av(\rv)$ to find the ground state densities.  But we did not do exactly this.  Rather, following the example of Kohn and Sham, who betrayed the spirit of their  theory before the rooster's crow, we resorted to the indignity of representing the non-interacting kinetic energy in terms of Kohn-Sham orbitals.  How else could one get a theory that was exact, at least,  in the noninteracting limit?  Thus, we wrote the full energy functional as
\be\label{CDFTFunctional}
E_{V,\Av}[n,\jv_p]=\sum_i \int \psi^*_i[n,\jv_p](\rv)\left\{\frac{1}{2}\left[-i\nabla_\rv+\Av(\rv)\right]^2+V(\rv)\right\}\psi_i[n,\jv_p](\rv) d\rv+E_H[n]+E_{xc}[n,\jv_p]
\ee
with the sum running over occupied orbitals and the densities given by Eqs.~(\ref{ParamagneticCurrent}) and ~(\ref{Density}).
Notice that, at this stage, the Kohn-Sham orbitals, $\psi_i[n,\jv_p](\rv) $, are still functionals of $n$ and $\jv_p$, in accordance to the Hohenberg-Kohn theorem for noninteracting system -- so the form of the theory is saved.  But the truth of the matter is that the functional dependence of $\psi_i$ on $n$ and $\jv_p$ is unknown: therefore, in order to find the Kohn-Sham orbitals, we must effectively treat them as independent variables, and zero the first-order variation of the functional under the infinitesimal increment $\psi_i^* \to \delta\psi_i^*$.  This is very easy to do, because the functional dependence of $n$ and $\jv_p$ on the the Kohn-Sham orbitals is known from Eqs.~(\ref{ParamagneticCurrent}) and ~(\ref{Density}). And this is how the KS orbital sneak into the theory as independent, rather than dependent variables.

All this was very standard in DFT, but in our case there was  a major difficulty stemming from the fact that the paramagnetic current density is not invariant under a ``gauge transformation", which changes the vector potential from $\Av(\rv)$ to $\Av(\rv)+\nablav_\rv\chi(\rv)$ while simultaneously multiplying all the Kohn-Sham orbitals by the same position-dependent phase factor $e^{i\chi(\rv)}$.  This transformation is an exact symmetry of the system and leaves the ground state energy unchanged.  However, under such a transformation, the paramagnetic current density changes according to 
\be\label{GaugeTransformation}
\jv_p(\rv)\to \jv_p(\rv) +n(\rv)\nablav_\rv\chi(\rv)\,,
\ee
while it can be rigorously proved~\cite{VR1987} that the xc energy functional is gauge-invariant in the sense that
\be
E_{xc}[n,\jv_p]=E_{xc}[\jv_p(\rv) +n(\rv)\nablav_\rv\chi(\rv)]
\ee
for any smooth gauge function $\chi(\rv)$.
And so, in order to ensure this fundamental property  we stipulated that $E_{xc}$ is a functional of the gauge-invariant  {\it vorticity} 
\be\label{GaugeTransformation}
\nuv(\rv)=\nablav_\rv\times\frac{\jv_p(\rv)}{n(\rv)}\,,
\ee
rather than directly of the paramagnetic current. In other words, we posited
\be\label{ExcBar}
E_{xc}[n,\jv_p]=\bar E_{xc}[n,\nuv]\,.
\ee
In making this crucial choice we were influenced by two considerations: first, we believed that the xc functional must be a genuine functional of local densities, i.e., it must not depend explicitly on Kohn-Sham orbitals; second, we observed that the vorticity  reduces to a constant $\propto \Bv$   in a uniform electron gas subjected to a spatially uniform magnetic field $\Bv$: this makes $\nuv$ an ideal candidate for a local density approximation. 

Armed with Eq.~(\ref{ExcBar}) we worked out the stationary condition for the functional~(\ref{CDFTFunctional}) by setting the first-order variation caused by the infinitesimal increment $\psi_i^* \to \psi_i^*+ \delta\psi_i^*$ to zero and we arrived
at the Kohn-Sham equation
 \be \label{GKS2}
\left(\frac{1}{2}\left[(-i\nablav_\rv+\Av)^2 + \left\{-i\nablav_\rv -\frac{\jv_p}{n},\cdot \Av_{xc}\right\}\right]+ V+V_H+\bar V_{xc}\right)\psi_i(\rv)=\varepsilon_i \psi_i(\rv)\,,
\ee
where  the curly bracket  represents the symmetrized product (anticommutator) $\{A,B\}=AB+BA$, $V_H$ is the  Hartree potential,
\be\label{VXC}
\bar V_{xc}(\rv)\equiv \left.\frac{\delta \bar E_{xc}[n,\nuv]}{\delta n(\rv)}\right\vert_{\nuv}
\ee
is the gauge-invariant $xc$ scalar potential and 
\be\label{AXC}
\Av_{xc}(\rv)\equiv\left.\frac{1}{n(\rv)}\nablav_\rv\times \frac{\delta \bar E_{xc}[n,\nuv]}{\delta \nuv(\rv)}\right\vert_{n}
\ee
is the $xc$ vector potential, also gauge-invariant.  The Kohn-Sham eigenvalues $\varepsilon_i$ on the right hand side of the equation emerge as  Lagrange multipliers ensuring the normalization of the corresponding orbitals.   

Notice that this equation is manifestly gauge invariant, as the gauge-invariant xc vector potential couples to the {\it relative} momentum operator
$-i\nablav -\jv_p/n$ (or, equivalently, $-i\nablav +\Av-\jv/n$), that is to say, the kinetic momentum evaluated in a reference frame in which a small volume of the electron gas, traveling with velocity $\jv/n$ in the ``laboratory frame", is at rest.\footnote{In Ref.~\cite{VR1987} this  equation was written in an equivalent but non manifestly gauge-invariant form, which employed the  non gauge-invariant  xc scalar potential   $V_{xc}\equiv \left.\frac{\delta E_{xc}[n,\jv_p]}{\delta n}\right\vert_{\jv_p} =\bar V_{xc}-\frac{\jv_p}{n}\cdot\Av_{xc}$.}

The Kohn-Sham equation of CDFT determines the density and the paramagnetic current density via Eqs.~(\ref{ParamagneticCurrent}) and ~(\ref{Density}).  An important observation is that the physical current density is still given by $\jv=\jv_p+n\Av$, which differs from the ``Kohn-Sham current", $\jv_{KS}=\jv_p+n\Av+n\Av_{xc}$ -- the latter  following directly from the ``Kohn-Sham velocity", i.e., the commutator of the position operator with the Kohn-Sham Hamiltonian.   This was a cause of  concern, since the structure of the Kohn-Sham equation only guarantees the conservation of the Kohn-Sham current, i.e., $\nablav_\rv\cdot\jv_{KS}(\rv)=0$ in the ground state, but not, a priori, the conservation of the physical current.  What saves the day is the fact that $\nablav_\rv\cdot[n(\rv)\Av_{xc}(\rv)]=0$ as a consequence of Eq.~(\ref{AXC}), so that $\nablav_\rv\cdot\jv(\rv)=\nablav_\rv\cdot\jv_{KS}(\rv)=0$ even as $\jv(\rv)\neq \jv_{KS}(\rv)$! The satisfaction of this  constraint played a decisive  role in convincing us that we had  found the correct generalization of the Kohn-Sham equation for electronic systems in magnetic fields.

\section{Generalized CDFT}\label{GeneralizedCDFT}
Against this background  it is not difficult to see why the 2002 paper by Axel Becke~\cite{Becke2002} came to me as a shock.  His xc energy functional, based on a  model for the exchange-correlation hole of a hydrogenic atom,  depended directly on the current density, not on the vorticity, in apparent contradiction with Eq.~(\ref{ExcBar}). His paper made no reference to our work, nor should have done so, since the VR formulation of CDFT obviously had  played no role in this development.  There was no magnetic field and no vorticity in Axel's paper -- and yet there was a current, and gauge-invariance was respected. What was going on?

The answer is that by this time Axel and I were standing on opposite sides of an ideological divide.  A paradigm shift had occurred during the nineties~\cite{Seidl96}. The use of KS orbitals, which was initially confined to the construction of the noninteracting kinetic energy functional had stealthily crept into the construction of the xc energy functional as well. At first, the shift was almost disguised: the orbitals were treated as  functionals of the density. This meant that the effective potentials -- now dubbed ``optimized effective potentials" -- were obtained through a rather cumbersome self-consistent procedure, involving the functional derivatives of the orbitals with respect to the density~\cite{OEP92,Kummel2003}. By the end of the nineties, however, the taboo against using orbitals in density functional approximations had all but disappeared and hybrid functionals were part of daily life. In a parallel development a new class of approximations, pioneered by people like John Dobson~\cite{Dobson92,Dobson93}, John Perdew~\cite{MGGA2003}, and Axel Becke himself~\cite{Becke88,BR89}, was emerging from a careful examination of the short range behavior of the exchange-correlation hole in atomic systems. These approximations expressed the xc energy as a functional of the local kinetic energy density -- the latter controlling the curvature of the xc hole at short distances -- and became the seed of what  today is widely known as ``meta-GGA"~\cite{MGGA2003}. The functional proposed by Becke in 2002 was precisely of this form.  It took the form
\be\label{ExcBecke}
E_{xc}[\{\psi_i\}]=\bar E_{xc}[n,\tilde\tau]
\ee
where $\tilde\tau(\rv)$  is the gauge-invariant kinetic energy density evaluated in a reference frame in which a volume element  of the electron gas is at rest.  Ignoring, for simplicity,  the spin we have
\ber\label{TauDef}
\tilde\tau(\rv) &=& \frac{1}{2}\sum_i\left\vert \nablav_\rv\psi_i(\rv)\right\vert^2 -\frac{|\jv_p(\rv)|^2}{2 n(\rv)}\nn\\
&=&\frac{1}{2}\sum_i \left\vert\left(-i\nablav_{\rv}-\frac{\jv_p}{n}\right)\psi_i(\rv)\right\vert^2\,.
\eer
Notice that the notation $\bar E_{xc}[n,\tilde\tau]$ should not be interpreted as saying that $\tilde \tau$ is a  basic variable on equal footing with the density: it simply means that $E_{xc}$ depends on the generalized Kohn-Sham orbitals via $\tilde \tau$, as well as via $n$. 

Observe the similarities and the differences between the VR functional~(\ref{ExcBar}) and the Becke functional~(\ref{ExcBecke}).  Both are manifestly gauge invariant, but the Becke form implicitly invokes, through  $\tilde\tau$,  the  current density, not the vorticity. 
Crucially, the Becke form adopts the orbitals as the basic variable -- the orbitals that determine the particle density $n$, the current density $\jv_p$, and the kinetic energy density $\tilde\tau$ -- the latter being extraneous to the VR formulation of CDFT.  

It is important to appreciate that the appearance of the current density in Eq.~(\ref{TauDef}) {\it has nothing to do with the presence or absence of an external magnetic field}: it is an inevitable consequence of the way complex orbitals enter the curvature of the exchange-correlation hole at short separations. Indeed, from the second line of Eq.~(\ref{TauDef}) we see that the inclusion of a magnetic vector potential does not change {\it the form} of $\tilde\tau$:
\be\label{TauDef2}
-i\nablav_{\rv}-\frac{\jv_p}{n} = -i\nablav_{\rv}+\Av-\frac{\jv}{n}\,.
\ee
This is not saying, of course, that the value of $\tilde\tau$ does not depend on the magnetic field: it does, but that dependence is entirely contained in the KS orbitals, which are now being treated (in stark contrast to the VR theory)  as a basic variable of the xc energy functional. 

Is this acceptable?  Well,  that depends on your ideological standing. It was clearly acceptable to Becke, who went on to show that his current-dependent xc functional could greatly improve the treatment of open-shell atoms.  By contrast,  vorticity-based functionals produced  no improvement or ran into significant numerical difficulties in atomic systems~\cite{ZhuTrickey2006}.

I do not know why neither Becke nor the  nearly contemporary proponents of this kind of approximations (I am thinking in particular of the work by the late Jianmin Tao and John Perdew\cite{Tao2005,TaoPerdew2005}, about which more will be said later) felt the need to work out (let alone to solve)  the generalized Kohn-Sham equation that follows from Eq.~(\ref{ExcBecke}).  If they did, they did not include it in their published work.  Apparently, they were satisfied with a non self-consistent solution of the problem. At any rate, the derivation of the self-consistent equations is quite a straightforward exercise -- all one has to do is to plug  the $E_{xc}$ functional of the form ~(\ref{ExcBecke}) in the energy functional of Eq.~(\ref{CDFTFunctional}) and carry out the minimization with respect to the orbitals, which now appear  not only in the noninteracting kinetic energy, but also in the xc energy functional.   But to do this, one must really embrace the generalized Kohn-Sham philosophy, and treat the orbitals as independent variables, not only in the noninteracting kinetic energy but also in the xc energy! 
The result is very simple and significantly different from the standard Kohn-Sham equation of CDFT. It can be written as follows:
%
\ber\label{GKS1}
&&\left\{\frac{1}{2}\left[\left(-i\nablav_\rv+\Av\right)^2+\left(-i\nablav_\rv-\frac{\jv_p(\rv)}{n(\rv)}\right)\frac{1}{m_{xc}(\rv)}\cdot\left(-i\nablav_\rv-\frac{\jv_p(\rv)}{n(\rv)}\right)\right]\right.\nn\\
&+&\left.V(\rv)+V_H(\rv)+\bar V_{xc}(\rv)\right\}\psi_i(\rv)= \varepsilon_i\psi_i(\rv)
\eer
where
\be
\bar V_{xc}(\rv)\equiv\left.\frac{\delta \bar E_{xc}[n, \tilde\tau]}{\delta n(\rv)}\right\vert_{\tilde\tau}\,,~~~~~~\frac{1}{m_{xc}(\rv)}\equiv\left.\frac{\delta \bar E_{xc}[n, \tilde\tau]}{\delta  \tilde\tau(\rv)}\right\vert_{n}\,.
\ee


Comparing Eqs.~(\ref{GKS1}) and (\ref{GKS2})  we observe a major difference:  the xc vector potential term (i.e., the second term of Eq.~(\ref{GKS2}), which depends {\it linearly} on  relative momentum)  has disappeared; in its place, a position-dependent  xc ``inverse effective mass" term (with $\frac{1}{m_{xc}(\rv)}$   the derivative of the xc energy with respect to the gauge-invariant kinetic energy density)  has emerged.  This term depends {\it quadratically} on relative momentum and  has no analogue in the  Kohn-Sham equation of VR~\cite{VR1987}. Ref.~\cite{Desmarais2026} (supplementary material, Eqs. (S41) and (S42)) shows that Eq.~(\ref{GKS1}) can be cast in a form similar to Eq.~(\ref{GKS2}) by defining a new xc vector potential $\Av_{xc}^{\tilde\tau} \equiv \left. \frac{\delta E_{xc}[n,\tilde\tau]}{\delta\jv_p}\right\vert_n = \left. \frac{\delta E_{xc}[n,\tilde\tau]}{\delta\tilde\tau}\right\vert_n  \left. \frac{\delta \tilde\tau }{\delta \jv_p}\right\vert_n= -\frac{\jv_p}{n m_{xc}}$.\footnote{The equations of Ref.~\cite{Desmarais2026} were written for the more general case of spin-current density functional theory. The result of interest here is obtained by dropping the spin variables and also ignoring the dependence of $E_{xc}$ on  gradients of the density.}  The price to be paid for this is that the xc scalar potential must  be modified by adding a mass term $\frac{1}{2} (-i\nablav_\rv)\frac{1}{m_{xc}(\rv)}(-i\nablav_\rv)$, which hardly fits our intuitive picture of what an effective scalar potential should look like. An equation equivalent to Eq.~(\ref{GKS1}) was first worked out by John Dobson -- see Eq. (26) of Ref.~\cite{Dobson92} -- but he did not seem to realize that the ``xc vector potential terms" could be absorbed in the quadratic $xc$ mass correction to the kinetic energy.

One might fear that the  mass term in the GKS equation~(\ref{GKS1}) will ruin the calculation of the current density, but this is not the case.  The GKS   velocity, computed as the commutator of the position operator with the GKS Hamiltonian, is
\ber
\hat \vv_{GKS}&=&\hat \vv+\frac{1}{2}\left\{-i\hat \nablav_\rv- \frac{\jv_p(\hat \rv)}{n(\hat \rv)},\frac{1}{m_{xc}(\hat\rv)}\right\} \nn\\
&=& \hat \vv+\frac{1}{2}\left\{-i\hat \nablav_\rv,\frac{1}{m_{xc}(\hat\rv)} \right\}- \frac{\jv_p(\hat\rv)}{n(\hat\rv)m_{xc}(\hat\rv)}\,,
\eer
where $\hat \vv$ is the physical velocity operator $-i\hat \nablav_\rv+\Av(\hat \rv)$.  The hat on  $\nablav_\rv$ emphasizes its operator character (i.e. $-i\hat \nablav_\rv$ is the momentum operator).  
With this, the GKS current density can be computed as
\ber\label{JGKS}
\jv_{GKS}(\rv)&=&\sum_i \Re e  [\psi^*_i(\rv) \hat \vv_{GKS}\psi_i (\rv)] \nn\\
&=&\jv(\rv)+\frac{1}{m_{xc}(\rv)}\left\{\frac{1}{2}\sum_i \Re e  [\psi^*_i(\rv) (-i\nablav_\rv)\psi_i(\rv)] -\frac{\jv_p(\rv)}{n(\rv)}\sum_i|\psi_i(\rv)|^2\right\}\,.
\eer
The quantity within the curly brackets vanishes (see Eq.~(\ref{ParamagneticCurrent}))
and we are left with the elegant  result 
\be
\jv_{GKS}(\rv)=\jv(\rv)\,,
\ee
i.e., the GKS current density coincides with the physical current density -- a distinct formal improvement upon the VR theory!

%
%

\section{Peaceful coexistence}\label{PeacefulCoexistence}
But why should we have to choose between the ``orthodox" VR formulation and the GKS formulation of CDFT?
The two formulations can be  combined in a single broader scheme.  All we have to do is posit an xc energy functional of the form
\be\label{Exc3}
E_{xc}[n,\jv_p]=\bar E_{xc}[n,\nuv,\tilde\tau]\,,
\ee
which depends both on the rest-frame kinetic energy density $\tilde\tau$ -- a functional of the orbitals independent of the magnetic field, as we have seen -- and on the vorticity imparted by the magnetic field.

This approach was pioneered by Tao and Perdew~\cite{Tao2005,TaoPerdew2005} (with less satisfactory results than Becke's), but great care must be exerted to avoid double-counting the contribution of the current density to the xc energy.  For weak magnetic field a natural approximation coming from perturbation theory is~\cite{TaoPerdew2005}
\be\label{Exc4}
\bar E_{xc}[n,\nuv,\tilde\tau]\simeq \bar E_{xc}[n,0,\tilde\tau]-\frac{1}{2}\int d\rv \chi_{L,xc}[n(\rv)] |\nuv(\rv)|^2 d\rv
\ee
where the first term on the right hand side is the meta-GGA xc energy functional for zero magnetic field,  $\chi_{L,xc}[n]$ is the difference between the interacting and the noninteracting orbital magnetic susceptibilities of the uniform electron gas of density $n$~\cite{VR1987}.  But the {\it value} of the kinetic energy density $\tilde\tau$, computed self-consistently from Kohn-Sham orbitals, is affected by the magnetic field,  so that the value of the first term of Eq.~(\ref{Exc3}) also depends on magnetic field, apparently contradicting the requirement that the magnetic field dependence of the xc energy come entirely from the second term of Eq.~(\ref{Exc4}).

To correct this potential double-counting Tao and Perdew suggested  that $\tilde\tau$ in the first term of Eq.~(\ref{Exc4}) should be modified by subtracting the vorticity contribution to the noninteracting kinetic energy in the presence of a magnetic field (see Eqs.(20) and (27) of Ref.~\cite{TaoPerdew2005}).  In this way, they claim, the approximate xc functional recovers the correct perturbative value in the uniform electron gas limit.

The essential point for the present discussion is that a hybrid meta-GGA+$\nuv$  functional of the form ~(\ref{Exc3})  can replace  $E_{xc}[n,\jv_p]$ in  the total energy functional~(\ref{CDFTFunctional}), and the latter can be minimized with respect to the generalized KS orbitals, which determine the three variables $n$,$\nuv$, and $\tilde\tau$ -- not the other way around!  The calculation is straightforward and we are led to the following GKS equation
\ber\label{GKS3}
&&\left\{\frac{1}{2}\left[(-i\nablav+\Av)^2 + \left\{-i\nablav-\frac{\jv_p}{n}, \cdot\Av_{xc}\right\}+\left(-i\nablav_\rv-\frac{\jv_p(\rv)}{n(\rv)}\right)\frac{1}{m_{xc}(\rv)}\cdot\left(-i\nablav_\rv-\frac{\jv_p(\rv)}{n(\rv)}\right)\right]\right.\nn\\
&+&\left.V(\rv)+V_H(\rv)+\bar V_{xc}(\rv)\right\}\psi_i(\rv)= \varepsilon_i\psi_i(\rv)
\eer
with 
\be\label{VXC-AXC}
\bar V_{xc}(\rv)\equiv\left.\frac{\delta \bar E_{xc}[n, \nuv, \tilde\tau]}{\delta n(\rv)}\right\vert_{\nuv,\tilde\tau}\,,~~~~~~\Av_{xc}(\rv)\equiv\left.\frac{1}{n(\rv)}\nablav_\rv\times \frac{\delta \bar E_{xc}[n,\nuv,\tilde\tau]}{\delta \nuv(\rv)}\right\vert_{n,\tilde\tau}
\ee
and
\be
\frac{1}{m_{xc}(\rv)}\equiv\left.\frac{\delta \bar E_{xc}[n, \nuv,\tilde\tau]}{\delta  \tilde\tau(\rv)}\right\vert_{n,\nuv}\,.
\ee
In this formulation, the xc vector potential of the VR theory peacefully coexists with the ``xc mass" term stemming from the explicit inclusion of orbitals in the kinetic energy density.  Just as in the VR theory the paramagnetic current is given directly by the GKS orbitals, using the standard formula (\ref{ParamagneticCurrent}). The  physical current is obtained by adding $n(\rv)\Av(\rv)$.

To the best of my knowledge, Eq.~(\ref{GKS3}), and even Eq.~(\ref{GKS2}) have never been tested  in physical systems.
I hope that this paper will inspire someone to do such a testing in a not too distant future.  

\section{Concluding remarks}\label{ConcludingRemarks}

In his humorous reconstruction of the evolution of DFT  Peter Gill\cite{Gill2002} casts Axel Becke as an ``eminent Canadian surgeon" who attempts to cure DFT of  ``advanced hyperparametric disorder" and thus ensures her immortality.  The surgeon's proposal was ``an abhorrent alliance with her brother (wave function theory) and the creation of a grisly hybrid" -- a concept that was ``as ghastly as it was irresistible".  These words (minus the grisly and the ghastly, of course)  capture the essence of Axel's  contribution to DFT.  Without his  ``heretic" contribution, which broke the Kohn-Sham orthodoxy, it is difficult to see how DFT would  have achieved its spectacular success, including the Nobel Prize in Chemistry.
As for myself, I must admit that I was unable to break free of the  ideological constraints that inspired my youthful enthusiasm for DFT, foremost among them the belief that the xc functional should be a universal functional of local densities, not orbitals.  This probably slowed down for a while the progress of CDFT, but the field has finally caught up with the times as one can see from the recent flurry of papers on current-spin-density functional theory (CSDF)~\cite{Desmarais2024,Desmarais2024-1,Desmarais2024-2,Desmarais2025,Desmarais2026} and the promising results that have been obtained therein.

\begin{acknowledgements}
    I thank Stefano Pittalis, Florian Eich and Jacques Desmarais for introducing me to the generalized Kohn Sham theory. This work was supported by the Ministry of Education, Singapore, under its funding for the Research Centre of Excellence award to the Institute for Functional Intelligent Materials. Project No. EDUNC-33-18-279-V12-IFIM.
\end{acknowledgements}

\bibliographystyle{apsrev4-2-titles}
\bibliography{CDFT-Orbital}

\end{document}